# ROBUST ADAPTIVE COMPENSATION OF EXTERNAL DISTURBANCES FOR MULTI-CHANNEL LINEAR SYSTEMS

V. H. Bui*, A. A Margun

*ITMO University, 197101, St. Petersburg, Russia*
*\*buinguyenkhanh201095@gmail.com*

**Abstract.** This paper proposes a new algorithm for compensating external disturbances for class of multi-channel linear systems. The solution to this problem is based on the use of the internal model principle and the extended error adaptation algorithm. It is assumed that the disturbance is the output of an autonomous linear generator with unknown parameters. At the first stage, a full-order observer with unknown input signals (Unknown Input Observer - UIO) is synthesized to solve the problem of estimating the state vector of this plant. Then a new observer of external disturbance is formed on the basis of state vector estimations. At the last stage, based on the new observer's estimations, a system with an extended state vector is formed for which a regulator providing compensation of disturbance is constructed. The performance of the obtained results is confirmed using computer simulation in MATLAB Simulink.

*Acknowledgements: The study was supported by the Ministry of Science and Higher Education of the Russian Federation, state assignment No. 2019-0898*

## INTRODUCTION

This paper considers the problem of compensation of external disturbances for a class multi-channel systems for the case of stationary and bounded amplitudes.

The problem of compensation of external disturbances is one of the fundamental and actual problems of the automatic control theory [1-3]. One of the widely used approaches is based on the internal model principle [4-6], where the external disturbances are described by the output of an autonomous linear generator. In works [7-10] the method of identification of parameters of external disturbances (phase, frequency and amplitude) is presented, then on the basis of obtained estimations the control law is synthesized. Classical methods of parameter identification are considered in [11].

An adaptive modification of the algorithm based on the principle of an internal model with discrete time is proposed in [12]. In [13-14], respectively, the method of dynamic regressor extension and mixing to estimate the unknown parameters of the dynamical system and the method of frequency estimation using sliding modes were studied. The advantage of this approach is that the operation of the identifier is independent of the regulator, which allows the application of various control and compensation methods. On the other hand, this approach has a significant limitation - the necessity to provide the condition of the regressor persistent excitation [15].

Another approach is the direct compensation of external disturbances, which solves the problem of the regressor persistent excitation [16-17]. For this case, a special observer of external disturbances is constructed using vectors of state variables or output signals. Based on the estimates of this observer, a direct controller is constructed to provide the desired properties of the closed-loop system. This method is effective and widely used for a class of single-channel systems. But the application to a class of multi-channel system with unknown state vector is still an open question.

In this paper we propose a new method of direct compensation of external disturbances, applicable to a class of multi-channel systems with an unknown state vector. A special observer is constructed to estimate the variables of the state vector of the plant. Then, based on the estimates of this observer, the observer of external disturbances is constructed. At the last stage, based on the estimates of disturbances, a control law is constructed to ensure the asymptotic convergence of the output signals to zero.

The paper is organized as follows: Section I is an brief of the problem. In section II we present mathematical problem statement and some assumptions. Parameterization of external disturbances is presented in section III. In section IV the full-order state observer is constructed. An observer an external disturbance observer is constructed in section V. Synthesis of the control law and adaptation algorithm are presented in section VI. The simulation results in Matlab are presented in section VII. Finally, in section VIII conclusions are presented.

To demonstrate the performance of the proposed method we conduct simulation in Matlab Simulink. The results of the simulation show that the goal of the work has been achieved: the boundedness of all signals in a closed-loop system, the tendency of the output signal of the system to zero when $t \to \infty$.

**PROBLEM STATEMENT**

Consider the class of linear stable disturbed control objects of the form:
$$\begin{cases} \dot{x}(t) = Ax(t) + Bu(t) + Ef(t), \\ y(t) = C^T x(t), \end{cases} \quad (1)$$

where $x(t) \in \mathbb{R}^n$ is an unmeasured state vector of the plant; $u(t) \in \mathbb{R}^\alpha$ is a control signal vector; $y(t) \in \mathbb{R}^\beta$ is a measurable output vector; $A, B, C, E$ are known constant matrices with an appropriate dimension; $f(t) \in \mathbb{R}^\gamma$ is unmeasured bounded external disturbance, where \gamma is such that \dim{\mathbf{Ef}(t)=n}.

This means that external disturbances affect the system arbitrarily. Just as many different disturbances can act on a single channel, so each disturbance can act on only one channel. Each perturbation $f_i(t)$ is a multiharmonic signal with unknown parameters:
$$f(t) = \sum_{j=0}^{p} R_j \sin(\omega_j t + \phi_j) + R_{0j},$$
where $i = \overline{1, \gamma}$; $R_j$ are unknown amplitudes; $\omega_j$ are frequencies; $\phi_j$ are phases and $R_{0j}$ are bias.

In this paper the following assumptions are accepted:

**Assumption 1.** The dimensionality of the disturbance and the maximum harmonic are known.

**Assumption 2.** Matrix $C$ has a full row rank and matrix $B$ has a full column rank.

**Assumption 3.** The pairs of known constant matrices $(A, B)$ and $(A, C)$ are controllable and observable respectively. The matrix $A$ is Hurwitz.

**Assumption 4.** The external disturbance vector $f(t)$ is bounded and can be represented as the output of a linear autonomous generator [19]:
$$\begin{cases} \dot{z}(t) = \Gamma z(t) \\ f(t) = h^T z(t), \end{cases}$$

where the parameters $\Gamma$, $h^T$ are unknown.

**Assumption 5.** The pair $(\Gamma, h^T)$ is fully observable and the eigenvalues of the $\Gamma$ lie on the imaginary axis.

If the external disturbance $f(t)$ is a multiharmonic signal then assumption 4 is always satisfied. But we must emphasize that from assumption 1 we know the dimensionality of the disturbance and the maximum harmonic, it follows that the dimensionality $q$ of the generator in assumption 5 is known. Assumption 2 guarantees a multi-channel system with a given number of inputs and outputs. Assumption 3 that the system is stable, since the paper focuses on compensating for external disturbances. Taking into account assumptions 1, 4, 5 the external disturbance can be consider as the output of a linear autonomous generator with unknown parameters but with a known limited number of harmonics.

The goal of this paper is the following: we need to construct a control law $u(t)$ that ensures that all signals in a closed system are bounded and that the output signal $y(t)$ tends to zero when time tends to infinity:

$$\lim_{t \to \infty} \|y(t)\| = 0$$

A schematic of the closed system of the proposed approach is shown in figure 1.

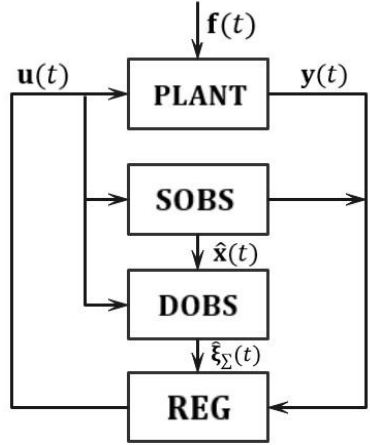

Figture 1. The structure of the closed-loop system scheme (SOBS is a full-order state observer, DOBS is an observer of external disturbances, $\hat{x}(t)$ is an estimate of the state vector, $\hat{\xi}_\sum(t)$ is an estimate of the regressor vector $\xi(t)$).

**PARAMETERIZATION OF EXTERNAL DISTURBANCES**

At the first stage the external disturbances are described by the output of an autonomous linear generator [2, 10, 16]:

$$\begin{cases} \dot{\xi}_\Sigma(t) = G_\Sigma \xi_\Sigma(t) + L_\Sigma f(t) \\ f(t) = \theta_\Sigma^T \xi_\Sigma(t) \end{cases} \quad (2)$$

where $\xi_\Sigma(t) = \begin{bmatrix} \xi_1 \\ \xi_2 \\ \vdots \\ \xi_q \end{bmatrix} \in \mathbb{R}^q$ is regressor, $G_\Sigma = \begin{bmatrix} G_1 & 0 & 0 & 0 \\ 0 & G_2 & 0 & 0 \\ 0 & 0 & \ddots & 0 \\ 0 & 0 & 0 & G_\gamma \end{bmatrix}$, $G_i$ are Hurwitz matrices;

$L_\Sigma = \begin{bmatrix} L_1 & 0 & 0 & 0 \\ 0 & L_2 & 0 & 0 \\ 0 & 0 & \ddots & 0 \\ 0 & 0 & 0 & L_\gamma \end{bmatrix}$, $L_i$ are constant vectors; $\theta_\Sigma^T \in \mathbb{R}^{\gamma \times q}$ is a vector of unknown constant parameters depending on the disturbance parameters. Pairs $(G_{i}, L_{i})$ are arbitrarily chosen such that each pair $(G_{i}, L_{i})$ is fully controllable.

## CONSTRUCTION OF A FULL-ORDER PLANT STATE OBSERVER

Since the values of the system state variables are unknown, it is necessary to form a full-order state observer to construct an external disturbance observer.

The structure of a full-order observer is described in [18]:
$$\begin{cases} \dot{w}(t) = Mw(t) + TBu(t) + Ky(t) \\ \hat{x}(t) = w(t) + Ny(t) \end{cases} \quad (3)$$

Where $w(t) \in \mathbb{R}^n$ is the state vector of the full-order observer; ; $\hat{x}(t) \in \mathbb{R}^n$ is state vector estimate; $M, T, K, N$ are the constants of the observer's matrices, chosen so as to satisfy the equations:
$$\begin{cases} (NC - I)E = 0 \\ T = I - NC \\ M = A - NCA - K_1 C \\ K_2 = MN \\ K = K_1 + K_2 \end{cases} \quad (4)$$

where $I$ is a unit matrix of appropriate dimensionality.

*Theorem 1:* Necessary and sufficient conditions for effective operation of the full-order state observer (3) for system (1) are [18]:

a) The rank of matrix $CE$ is equal to the rank of matrix $E$, i.e
$$rank(CE) = rank(E)$$

b) Matrix pair $(C, A)$ is a detectable pair.

where $\bar{A} = A - E[(CE)^T CE]^{-1}(CE)^T CA$

By introducing the state estimation error $e_x = x - \hat{x}$ and differentiating it with (1) in time, we obtain a dynamic model of observation error:
$$\dot{e}_x(t) = (A - NCA - K_1 C)e_x(t) + [M - (A - NCA - K_1 C)]w(t) + [K_2 - (A - NCA - K_1 C)N]y(t) + [T - (I - NC)]Bu(t) + (NC - I)Ef(t) \quad (5)$$

Substituting expression (4) into (5) we obtain:
$$\dot{e}_x(t) = Me_x(t)$$

The matrix $M$ can be computed such that $M \prec 0$, then $e(t)$ asymptotically converges to zero, i.e. $\hat{x}(t) \to x(t)$.

**Remark 1.** If the matrix $E$ does not have a full column rank, we can break the matrix $E$ into $E = \{E_1\}\{E_2\}$. Where $E_1$ has a full column rank and $E_{2}f(t)$ is considered a new external disturbance.

**Remark 2.** The matrices M, T, K, N, $K_1$, $K_2$ are determined by the following algorithm:
1) Check that the equality is fulfilled $rank(C^T E) = rank(E)$. If $rank(C^T E) \neq rank(E)$ exists the observer (3) for systems (1).
2) Calculate the matrices $N, T, A_1$
$N = E[(CE)^T CE]^{-1}(CE)^T; T = I - NC; A_1 = TA$
3) Check the observability of a pair matrices $(C, A_1)$, if the pair $(C, A_1)$ is observable, then $(C, A)$ is also observable. We can skip the observability check of the matrix pair $(C, A_1)$. The matrix $K\_1$ can be calculated using the pole placement method, then go to step 9.
4) 4) If $(C, A_1)$ is not observable, we must calculate the auxiliary matrix $P$

$$P = [p_1, \ldots p_{n_0}; p_{n_0+1}, \ldots, p_n]^T$$

Where $n_1 = rank(W_0)$, $W_0$ is the observability matrix of $(C, A_1)$; row vector $p_1^T, \ldots, p_{n_1}^T$ from $W_0$, together other $n - n_1$ row vector $p_{n_1+1}^T, \ldots, p_n^T$.

5) 5) Perform canonical decomposition observable on $(C, A_1)$:

$$PA_1 P^{-1} = \begin{bmatrix} A_{11} & 0 \\ A_{12} & A_{22} \end{bmatrix}, \ A_{11} \in \Box^{n_1 \times n_1}.$$

$$CP^{-1} = [C^* \ \ 0], \ C^* \in \Box^{m \times n_1}.$$

Where $n_1 = rank(W_0)$.

6) If the eigenvalues of $A_{22}$, $\lambda_i, i = \overline{1, n - n_1}$ lie in the left half-plane then $A$ is stable and otherwise observer does not exist.
7) Use $n_1$ desirable eigenvalues to construct $A_{11} - K_p^1 C^*$ using pole placement.
8) Compute $K_1 = P^{-1} K_p = P^{-1}[(K_p^1)^T \ \ (K_p^2)^T]^T$. Where $K_p^2$ is an arbitrary matrix of size $(n - n_1)m$.
9) Finally we compute the matrix:
$$M = A_1 - K_1 C, \ K = K_1 + K_2 = K_1 + MN.$$

## CONSTRUCTION OF AN EXTERNAL DISTURBANCE OBSERVER

Based on the estimates of the state vector, we form an observer of external disturbances [19]:

$$\begin{cases} \hat{\xi}_\Sigma(t) = \varphi_\Sigma + \mathbf{Q}_\Sigma \mathbf{x}(t) \\ \dot{\varphi}_\Sigma(t) = \mathbf{G}_\Sigma \varphi_\Sigma(t) + (\mathbf{G}_\Sigma \mathbf{Q}_\Sigma + \mathbf{Q}_\Sigma \mathbf{A})\mathbf{x}(t) - \mathbf{Q}_\Sigma \mathbf{B} \mathbf{u}(t) \end{cases} \quad (6)$$

where $\hat{\boldsymbol{\xi}}_\Sigma(t) = \begin{bmatrix} \hat{\boldsymbol{\xi}}_1 \\ \hat{\boldsymbol{\xi}}_2 \\ \vdots \\ \hat{\boldsymbol{\xi}}_q \end{bmatrix}$, $\hat{\boldsymbol{\xi}}_\Sigma(t) \in \mathbb{R}^q$ is vector estimate $\boldsymbol{\xi}_\Sigma(t)$; $\boldsymbol{\varphi}_\Sigma(t) = \begin{bmatrix} \boldsymbol{\varphi}_1 \\ \boldsymbol{\varphi}_2 \\ \vdots \\ \boldsymbol{\varphi}_q \end{bmatrix}$, $\boldsymbol{\varphi}_\Sigma \in \mathbb{R}^q$ is auxiliary vector of the observer; Matrix $\mathbf{Q}_\Sigma = \begin{bmatrix} \mathbf{Q}_1 \\ \mathbf{Q}_2 \\ \vdots \\ \mathbf{Q}_\gamma \end{bmatrix}$, $\mathbf{Q}_\Sigma \in \mathbb{R}^{q \times n}$ is satisfie:

$$\mathbf{Q}_i \mathbf{E} = \mathbf{L}_{0i}$$

where i=\overline{1,\gamma} is the number of the observer corresponding to the external disturbance and the matrix L_{0i}:

$$\mathbf{L}_{0i} = [\mathbf{0}_{qi}, \dots, \mathbf{0}_{qi}, \mathbf{L}_i, \mathbf{0}_{qi}, \dots, \mathbf{0}_{qi}]$$

contains vector $L_i$ as the i-th column, and $0_{qi}$ is the $q_i$-dimensional zero vector. This means that:

$$\mathbf{L}_\Sigma = \begin{bmatrix} \mathbf{L}_{01} \\ \mathbf{L}_{02} \\ \vdots \\ \mathbf{L}_{0\gamma} \end{bmatrix}$$

As a result, the external disturbance can be represented as:

$$\mathbf{f}(t) = \boldsymbol{\theta}_\Sigma^T \hat{\boldsymbol{\xi}}_\Sigma(t) + \boldsymbol{v}$$

where $\boldsymbol{\theta}_\Sigma^T = [\boldsymbol{\theta}_1, \boldsymbol{\theta}_2 \dots \boldsymbol{\theta}_\gamma]^T \in \mathbb{R}^{\gamma \times q}$; $\hat{\boldsymbol{\xi}}_\Sigma(t) = [\hat{\boldsymbol{\xi}}_1^T, \hat{\boldsymbol{\xi}}_2^T, \dots, \hat{\boldsymbol{\xi}}_\gamma^T]^T \in \mathbb{R}^q$; $\upsilon$ - экспоненциально затухающая функция.

## SYNTHESIS OF THE CONTROL LAW AND ADAPTATION ALGORITHM

To compensate for external disturbances, we construct a regulator based on the works [20, 21]. Let us change the coordinates of external disturbance into the coordinate frame of the plant state vector using the transformation matrix $M$. The parametric tracking error of plant state takes the form:

$$\mathbf{e}(t) = \mathbf{x}(t) - \mathbf{M}\boldsymbol{\xi}_\Sigma(t), \qquad (7)$$

By differentiating (7), taking into account (1) and (2), we obtain:

$$\dot{\mathbf{e}} = \mathbf{A}\mathbf{e} + [\mathbf{A}\mathbf{M} - \mathbf{M}(\mathbf{Q}_\Sigma + \mathbf{L}_\Sigma \boldsymbol{\theta}_\Sigma^T) + \mathbf{E}\boldsymbol{\theta}_\Sigma^T]\boldsymbol{\xi}_\Sigma(t) + \mathbf{B}u(t)$$

and the output signal:

$$\mathbf{y} = \mathbf{C}^T \mathbf{e} + \mathbf{C}^T \mathbf{M}\boldsymbol{\xi}_\Sigma(t)$$

Exist matrices $M$ and $\psi$ such that the system of equations

$$\begin{cases} \mathbf{A}\mathbf{M} - \mathbf{M}(\mathbf{Q}_\Sigma + \mathbf{L}_\Sigma \boldsymbol{\theta}_\Sigma^T) = \mathbf{B}\boldsymbol{\psi}_\Sigma^T - \mathbf{E}\boldsymbol{\theta}_\Sigma^T \\ \mathbf{C}^T \mathbf{M} = \mathbf{0} \end{cases}$$

has at least one solution and are called Francis or regulator equations [21]. The corresponding matrix $\boldsymbol{\psi}_\Sigma^T = [\boldsymbol{\psi}_1, \boldsymbol{\psi}_2 \dots \boldsymbol{\psi}_\alpha]^T \in \mathbb{R}^{\alpha \times q}$.

The dynamics of error model take the form:

$$\begin{cases} \dot{\mathbf{e}} = \mathbf{Ae} + \mathbf{B}(\boldsymbol{\psi}_{\Sigma}^{\mathrm{T}}\hat{\boldsymbol{\xi}}_{\Sigma}(t) + \mathbf{u}) + \upsilon \\ \mathbf{y} = \mathbf{C}^{\mathrm{T}}\mathbf{e} \end{cases},$$

where $\upsilon$ is an exponentially decaying function.

Thus, the control law can be constructed $\mathbf{u} = -\widehat{\boldsymbol{\psi}}_{\Sigma}^{\mathrm{T}}\hat{\boldsymbol{\xi}}_{\Sigma}$.

The resulting error model take the form of:
$$\begin{cases} \dot{\mathbf{e}} = \mathbf{Ae} + \mathbf{B}\widetilde{\boldsymbol{\psi}}_{\Sigma}^{\mathrm{T}}\hat{\boldsymbol{\xi}}_{\Sigma}(t) + \upsilon \\ \mathbf{y} = \mathbf{C}^{\mathrm{T}}\mathbf{e} \end{cases}, \tag{8}$$

The output vector of the system (8) can be rewritten in the form:
$$\mathbf{y} = \mathbf{W}(s)\left[\widetilde{\boldsymbol{\psi}}_{\Sigma}^{\mathrm{T}}\hat{\boldsymbol{\xi}}_{\Sigma}(t)\right] + \upsilon,$$

where $\mathbf{W}(s) = \mathbf{C}^{\mathrm{T}}(s\mathbf{I} - \mathbf{A})^{-1}\mathbf{B}$.

### A. Gradient Algorithm of Adaptation

Using the method of extended error, we obtain:
$$\bar{\mathbf{y}} = \mathbf{y} - \hat{\mathbf{y}},$$
$$\bar{\mathbf{y}} = \mathbf{y} - \mathbf{W}(s)\left[\hat{\boldsymbol{\xi}}_{\Sigma}^{\mathrm{T}}\right]\widehat{\boldsymbol{\psi}}_{\Sigma} - \mathbf{W}(s)[\mathbf{u}] = \mathbf{W}(s)\left[\hat{\boldsymbol{\xi}}_{\Sigma}\right]\widetilde{\boldsymbol{\psi}}_{\Sigma} + \upsilon, \tag{9}$$

where $\mathbf{W}(s)[\hat{\boldsymbol{\xi}}_{\Sigma}^{\mathrm{T}}] = \begin{bmatrix} W_{11}(s)[\hat{\boldsymbol{\xi}}_{\Sigma}^{\mathrm{T}}] & W_{12}(s)[\hat{\boldsymbol{\xi}}_{\Sigma}^{\mathrm{T}}] & \dots & W_{1\alpha}(s)[\hat{\boldsymbol{\xi}}_{\Sigma}^{\mathrm{T}}] \\ W_{21}(s)[\hat{\boldsymbol{\xi}}_{\Sigma}^{\mathrm{T}}] & W_{22}(s)[\hat{\boldsymbol{\xi}}_{\Sigma}^{\mathrm{T}}] & \dots & \vdots \\ \vdots & \vdots & \ddots & \vdots \\ W_{\beta 1}(s)[\hat{\boldsymbol{\xi}}_{\Sigma}^{\mathrm{T}}] & \dots & \dots & W_{\beta\alpha}(s)[\hat{\boldsymbol{\xi}}_{\Sigma}^{\mathrm{T}}] \end{bmatrix}$.

Note that $\boldsymbol{\psi}_{\Sigma}^{\mathrm{T}}$ is a vector then we can put it outside the brackets $\boldsymbol{\psi}_{\Sigma}^{\mathrm{T}}\mathbf{W}(s)[\hat{\boldsymbol{\xi}}_{\Sigma}]$. In this paper, $\boldsymbol{\psi}_{\Sigma}^{\mathrm{T}}$ is a matrix, so putting it outside the brackets is not allowed. $\mathbf{W}(s)[\boldsymbol{\psi}_{\Sigma}^{\mathrm{T}}\hat{\boldsymbol{\xi}}_{\Sigma}]$ can be represented as follows:
$$\mathbf{W}(s)[\widehat{\boldsymbol{\psi}}_{\Sigma}^{\mathrm{T}}\hat{\boldsymbol{\xi}}_{\Sigma}^{\mathrm{T}}] = [W_{i1}(s)[\hat{\boldsymbol{\xi}}_{\Sigma}^{\mathrm{T}}] + W_{i2}(s)[\hat{\boldsymbol{\xi}}_{\Sigma}^{\mathrm{T}}] + \cdots + W_{i\alpha}(s)[\hat{\boldsymbol{\xi}}_{\Sigma}^{\mathrm{T}}]]\widehat{\boldsymbol{\psi}}_{\Sigma}$$
i.e:
$$\mathbf{W}(s)[\widehat{\boldsymbol{\psi}}_{\Sigma}^{\mathrm{T}}\hat{\boldsymbol{\xi}}_{\Sigma}] = \mathbf{W}(s)[\hat{\boldsymbol{\xi}}_{\Sigma}^{\mathrm{T}}]\widehat{\boldsymbol{\psi}}_{\Sigma}$$

Based on equation (9), we construct a standard adaptation algorithm:
$$\dot{\widehat{\boldsymbol{\psi}}}_{\Sigma} = \gamma \mathbf{W}(\mathbf{s})[\hat{\boldsymbol{\xi}}_{\Sigma}^{\mathrm{T}}]\bar{\mathbf{y}}, \tag{10}$$

В [22] рассмотрены свойства стандартного адаптивного алгоритма. Он обеспечивает достаточно высокое время переходных процессов. Для улучшения сходимости всех сигналов в системе к нулю предлагается использовать альтернативные алгоритмы адаптации [23].

### B. Adaptation Algorithm With Memory Regressor Extension

From equation (9) we obtain:
$$\hat{\mathbf{y}} = \mathbf{y} - \mathbf{W}(s)[\mathbf{u}].$$

Take the transfer function L(s) on both sides we obtain:
$$\underbrace{\mathbf{H}(s)[\Delta\bar{\mathbf{y}}]}_{\mathbf{Y}} = \underbrace{\mathbf{H}(s)[\Delta\Delta^{\mathrm{T}}]}_{\Omega}\boldsymbol{\psi}_{\Sigma} \tag{11}$$

Where $\Delta = \mathbf{W}(s)[\hat{\boldsymbol{\xi}}_{\Sigma}]$, $\mathbf{H}(s) = \frac{1}{\alpha s + 1}$, $\alpha > 0$ - is asymptotically stable and minimal phase transfer function.

На основе уравнения (11) построим альтернативные алгоритмы адаптации [23]:

$$\dot{\widehat{\boldsymbol{\psi}}}_\Sigma = \gamma(\mathbf{Y} - \boldsymbol{\Omega}\widehat{\boldsymbol{\psi}}_\Sigma), \gamma > 0. \tag{12}$$

So, we can conclude that the proposed approach allows compensating external disturbances taking into account the above assumptions 1-5. By choosing the adaptation coefficient $\gamma>0$, the boundedness of all signals in the closed-loop system and the goal equality is ensured:

$$\lim_{t\to\infty}\|\mathbf{y}(t)\| = 0.$$

## SIMULATION RESULTS

To demonstrate the performance of the proposed method we conduct simulation in Matlab Simulink.

As an example, consider a third-order system:

$$\begin{cases} \dot{\mathbf{x}}(t) = \begin{bmatrix} -1 & 1 & 0 \\ 0 & 0 & 1 \\ -4 & -5 & -6 \end{bmatrix} \mathbf{x}(t) + \begin{bmatrix} 2 & 0 \\ 1 & 0 \\ -1 & 3 \end{bmatrix} \mathbf{u}(t) + \begin{bmatrix} -1 & 0 \\ 0 & 0 \\ -1 & 1 \end{bmatrix} \mathbf{f}(t) \\ \mathbf{y}(t) = \begin{bmatrix} 1 & 0 & 0 \\ 0 & 1 & 1 \end{bmatrix} \mathbf{x}(t) \end{cases},$$

with initial conditions $\mathbf{x}(t) = [1; 1; 0]$.

External disturbances $\mathbf{f}(t) = \begin{bmatrix} 5\sin(2t) \\ 4 + 7\sin(3t) \end{bmatrix}$ are described by the output of autonomous linear generators with matrices:

$$\mathbf{G}_1 = \begin{bmatrix} 0 & 1 \\ -3 & -4 \end{bmatrix}, \mathbf{L}_1 = \begin{bmatrix} 0 \\ 2 \end{bmatrix} \text{ и } \mathbf{G}_2 = \begin{bmatrix} 0 & 1 & 0 \\ 0 & 0 & 1 \\ -6 & -11 & -6 \end{bmatrix}, \mathbf{L}_2 = \begin{bmatrix} 0 \\ 0 \\ 6 \end{bmatrix}.$$

To construct a full-order state observer (3), we choose the matrix $\mathbf{K} = \begin{bmatrix} 0 & 0 \\ 0 & 1 \\ 0 & -1 \end{bmatrix}$, where

$$\mathbf{K}_1 = \begin{bmatrix} 3 & -5 \\ -1 & 5 \\ -3 & 7 \end{bmatrix}.$$

Construct an observer of external disturbances (6) with the following matrices:

$$\mathbf{Q}_1 = \begin{bmatrix} 0 & 0 & 0 \\ -2 & 0 & 0 \end{bmatrix}, \mathbf{Q}_2 = \begin{bmatrix} 0 & 0 & 0 \\ 0 & 0 & 0 \\ -6 & 0 & 6 \end{bmatrix}.$$

Figure 2 shows the transients when using the standard adaptation algorithm (10) with an adaptation gain $\gamma=5$: output signal vector $\mathbf{y}(t)$ (a); estimation errors of the state vector $\mathbf{e}_\mathbf{x}(t) = \mathbf{x}(t) - \hat{\mathbf{x}}(t)$ (b); estimates of the tunable parameter vector $\widehat{\boldsymbol{\psi}}_\Sigma$ (c); control signal vector $\mathbf{u}(t)$ (d).

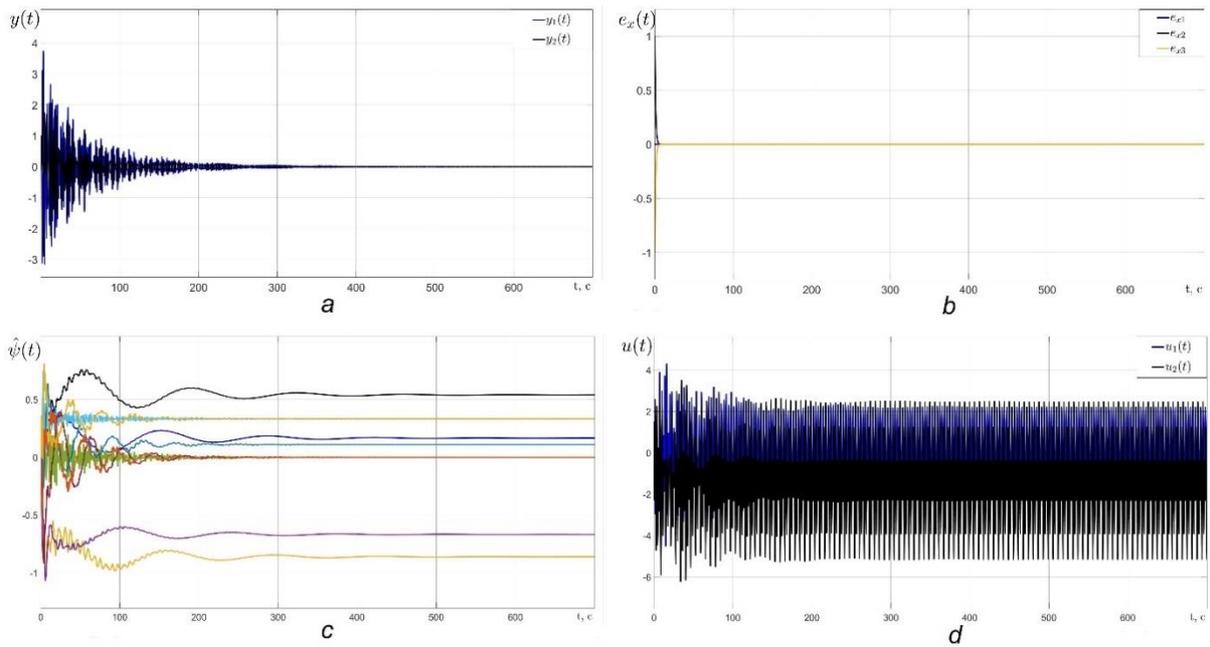

Figture 2. Graphs of transients of the standard adaptation algorithm (10) with adaptation coefficient $\gamma = 5$: output signal vector $\mathbf{y}(t)$ (a); estimation errors of the state vector $\mathbf{e}_\mathbf{x}(t) = \mathbf{x}(t) - \hat{\mathbf{x}}(t)$ (b); estimates of the tunable parameter vector $\hat{\boldsymbol{\psi}}_\Sigma$ (c); control signal vector $\mathbf{u}(t)$ (d).

Figure 3 shows the transients of the alternative adaptation algorithm (12) c $\mathbf{H}(s) = \frac{1}{s+1}$ with an adaptation gain $\gamma=25$: output signal vector $\mathbf{y}(t)$ (e); estimation errors of the state vector $\mathbf{e}_\mathbf{x}(t) = \mathbf{x}(t) - \hat{\mathbf{x}}(t)$ (f); estimates of the tunable parameter vector $\hat{\boldsymbol{\psi}}_\Sigma$ (g); control signal vector $\mathbf{u}(t)$ (h).

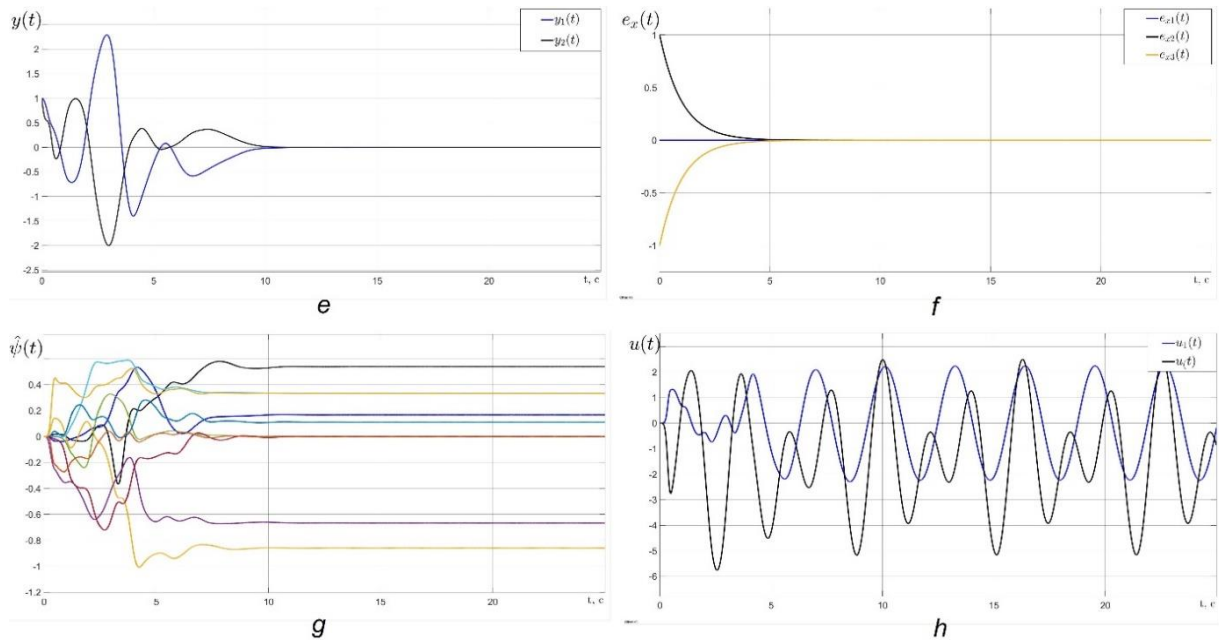

Figure 3. Graphs of the transients of the alternative adaptation algorithm (12) with the adaptation coefficient $\gamma=25$: output signal vector $\mathbf{y}(t)$ (e); estimation errors of the state vector $\mathbf{e_x}(t) = \mathbf{x}(t) - \hat{\mathbf{x}}(t)$ (f); estimates of the tunable parameter vector $\widehat{\boldsymbol{\psi}}_\Sigma$ (g); control signal vector $\mathbf{u}(t)$ (h).

The simulation results show that when using the standard adaptation algorithm (10), the transient time of the system is so long. When using the alternative adaptation algorithm (12), the transient time is decreased significantly. After 5 seconds of simulation, the state vector of the plant converges to the true value $x(t)$, which allows us to conclude that it works correctly. Analysis of the simulation results demonstrates that the proposed approach ensures that all signals in the closed system are bounded and the goal equality $\lim_{t\to\infty}\|\mathbf{y}(t)\| = 0$ with assumptions 1-5.

## CONCLUSION

In this paper we consider the problem of compensation of external unknown disturbances for a class of multi-channel linear systems. A new approach for the case of an unmeasurable state vector is proposed. A full-order state observer that ensures the convergence of the state vector to the true value is constructed. Simulation results demonstrate the performance of the proposed approach, as well as a reduction in the convergence time when using an alternative adaptation algorithm. In the future, the obtained solution can be extended to the class of linear and nonlinear time-delayed system.